\documentclass[twocolumn,aps,prd,showpacs,superscriptaddress,nofootinbib,amsmath,amssymb,floats,floatfix,showkeys,notitlepage,longbibliography]{revtex4-1}

\usepackage{comment}
\usepackage{graphicx}
\usepackage[]{mdframed}
\usepackage{subfigure}
\usepackage{palatino}
\usepackage[commandnameprefix=always]{changes}
\usepackage{hyperref}
\hypersetup{colorlinks=true,linkcolor=blue,urlcolor=blue,citecolor=blue}
\usepackage[toc,page]{appendix}
\usepackage[normalem]{ulem}

\usepackage{orcidlink}
\usepackage{lipsum}
\usepackage{graphicx}
\usepackage{subfigure}
\usepackage{palatino}
\usepackage{sans}
\usepackage{adjustbox}
\usepackage{latexsym}
\usepackage{amsmath}
\usepackage{amssymb}
\usepackage{amsfonts}
\usepackage{dcolumn}
\usepackage{bm}
\usepackage{tikz}
\usepackage{bigints}
\usepackage{array,tabularx,multirow,booktabs}
\usepackage[tracking=true]{microtype}
\SetTracking{}{500}
\SetTracking{encoding={*}, shape=sc}{40}
\UseRawInputEncoding 
\allowdisplaybreaks
\usepackage{adjustbox}
\usepackage{latexsym}
\usepackage{amsmath}
\usepackage{amssymb}
\usepackage{amsfonts}
\usepackage{dcolumn}
\usepackage{bm}
\usepackage{tikz}
\usepackage{bigints}
\usepackage{array,tabularx,multirow,booktabs}
\usepackage[tracking=true]{microtype}
\usepackage{color}
\UseRawInputEncoding 
\allowdisplaybreaks

\begin{document} 

\title{Formation of regular black hole from baryonic matter}

\author{Vitalii Vertogradov}
\email{vdvertogradov@gmail.com}
\affiliation{Physics department, Herzen state Pedagogical University of Russia,
48 Moika Emb., Saint Petersburg 191186, Russia} 
\affiliation{Center for Theoretical Physics, Khazar University, 41 Mehseti Street, Baku, AZ-1096, Azerbaijan.}
\affiliation{SPB branch of SAO RAS, 65 Pulkovskoe Rd, Saint Petersburg
196140, Russia}
\author{Ali \"Ovg\"un
}
\email{ali.ovgun@emu.edu.tr}
\affiliation{Physics Department, Eastern Mediterranean
University, Famagusta, 99628 North Cyprus, via Mersin 10, Turkiye.}

\author{Daniil Shatov}
\email{shatov.ddb@mail.ru}
\affiliation{Physics department, Herzen state Pedagogical University of Russia,
48 Moika Emb., Saint Petersburg 191186, Russia}

\date{\today}

\begin{abstract}
\textcolor{black}{We present a family of exact, singularity-free solutions describing the collapse of baryonic matter characterized by a barotropic equation of state whose coefficient $\alpha(r,v)$ varies in both radius and time. By matching these interior solutions to the Husain exterior metric, we obtain a self-consistent, dynamical spacetime representing a regular black hole. Although the pressure profile of our models grows with radius and eventually violates the dominant energy condition beyond a critical surface-necessitating an external junction to ensure a globally well-defined spacetime-the interior solution remains non-singular throughout the collapse. We further analyze the optical properties of these regular black holes and find that both the photon sphere radius and the corresponding shadow radius increase monotonically as the local equation of state parameter $\alpha$ is raised. Moreover, the matching interface between the interior and exterior metrics naturally suggests a phase transition in the collapsing fluid, which can postpone the formation of an apparent horizon.  Taken together, our results not only highlight novel physical features of horizon formation in regular collapse models but also identify characteristic shadow signatures that could be tested by future observations.}

\end{abstract}


\maketitle

\section{Introduction}

Penrose's singularity theorems demonstrate that apparent horizon formation inevitably leads to spacetime singularities \cite{Penrose:1964wq,Penrose:1969pc}, highlighting a fundamental limitation of general relativity in describing physics at extreme gravitational scales \cite{Hawking:1970zqf}. The Event Horizon Telescope's groundbreaking observations of supermassive black holes in M87 and Sagittarius A* have transformed these theoretical objects into observable reality, bringing renewed urgency to the singularity problem\cite{EventHorizonTelescope:2019dse,EventHorizonTelescope:2022wkp,Koyuncu:2014nga,Donmez:2022dze,Donmez:2023egk}. However, the singularity theorems assume the strong energy condition - that gravity remains universally attractive. This assumption can be violated by exotic matter states, with dark energy serving as a cosmic-scale example of repulsive gravity. Such violations could potentially resolve the singularity problem by preventing the formation of infinite density states.

The concept that extremely dense matter might transition into a vacuum state resembling a de Sitter core was first proposed independently by Gliner \cite{Gliner1966JETP}  and Sakharov in 1966 \cite{Sakharov1966JETP}. This groundbreaking insight laid the foundation for understanding potential mechanisms to avoid singularities in black holes. Building on these ideas, Bardeen made a significant advance in 1968 by constructing the first explicit model of a non-singular black hole \cite{Bardeen1968qtr}. However, a crucial question remained unresolved for nearly three decades: what type of matter could physically support such a regular center? This theoretical puzzle was finally addressed when Ayon-Beato and Garcia demonstrated that nonlinear electrodynamics could serve as the source for the Bardeen black hole, providing a concrete physical mechanism for singularity avoidance \cite{Ayon-Beato:1998hmi,Ayon-Beato:1999qin}.

Nonlinear electrodynamics as a mechanism for regular black hole formation faces fundamental challenges. The primary limitation stems from the non-uniqueness of the theory - there exists a vast family of nonlinear electromagnetic theories capable of generating regular centers, with no clear physical principle to select among them. This theoretical redundancy suggests that nonlinear electrodynamics may not provide the most fundamental explanation for singularity avoidance \cite{Hayward:2005gi, Verbin:2024ewl,Sorokin:2021tge,Balart:2023odm,Okyay:2021nnh}.The existence of regular centers in black hole solutions has been established only under specific conditions: the presence of magnetic monopoles and the complete absence of electric charge. This constraint significantly restricts the physical applicability of these models, particularly given that magnetic monopoles remain unobserved in nature ~\cite{Bronnikov:2000vy,Bronnikov:2006fu,Bronnikov:2017tnz}. Regular black holes supported by nonlinear electrodynamics rely on charge as a key parameter for regularization. However, real astrophysical black holes are generally considered to be electrically neutral. Even if a regular center were to form, it would likely be transient, as a singularity would inevitably develop due to the charged Penrose process ~\cite{Denardo:1973pyo, Zaslavskii:2020crn, Vertogradov:2022eeq} or through the accretion of matter onto the black hole.

Another fundamental challenge is understanding the formation of regular black holes. While a regular center necessitates the presence of exotic matter, regular stars are composed of ordinary baryonic matter. Consequently, during gravitational collapse, ordinary matter would have to undergo a transformation into an exotic form capable of preventing singularity formation. A recent model \cite{Vertogradov:2025yto} describes the gravitational collapse of dust and radiation, where dust transitions into radiation—a process that intensifies near the center—potentially facilitating the formation of a regular core. Additionally, previous studies have explored the gravitational collapse of dust leading to the emergence of regular black holes \cite{Bonanno:2023rzk}. Numerous studies have been dedicated to investigating the properties of static and stationary regular black holes ~\cite{Ovalle:2024wtv,Ghosh:2022gka,Capozziello:2024ucm,Misyura:2024fho,Konoplich:1999qq, Khlopov:1999ys, Khlopov:2008qy, Nicolini:2005vd,Vagnozzi:2022moj,Belotsky:2014kca, Dymnikova:2015yma, Dymnikova:1992ux,Bambi:2013ufa, Dymnikova:2001fb, Molina:2021hgx,Azreg-Ainou:2014nra, Mazharimousavi:2019jja,Halilsoy:2013iza,Santos:2024vby, Carballo-Rubio:2025fnc, Sueto:2023ztw, Ovgun:2019wej,Ovgun:2024zmt,AraujoFilho:2024lsi,Heidari:2024bkm,Filho:2024zxx,Bokulic:2023afx,Abdujabbarov:2016hnw, Feng:2023pfq, Allahyari:2019jqz,Fernando:2012yw,Kleihaus:2007vf,Kleihaus:1998qc,Toshmatov:2017zpr,Stuchlik:2019uvf,Sharif:2022oym,Sharif:2010pj,Hod:2024aen,Zaslavskii:2010qz, Kumar:2019pjp,Kumar:2020yem,KumarWalia:2022aop,Murk:2023rwl, Ghosh:2014hea,Smith:1998qx} (see~\cite{Ansoldi:2008jw, Lan:2023cvz} for a comprehensive review and references therein). However, comparatively fewer studies have focused on the problem of their formation \cite{Lan:2023cvz, Hayward:2005gi,Zhang:2014bea, Shojai:2022pdq, Mosani:2023awd, Bueno:2024eig, Bueno:2024zsx}.

In this work, we investigate the gravitational collapse of baryonic matter characterized by a dynamical equation of state (EoS) with time- and radius- 
 dependent coefficients. Through rigorous analysis of Einstein's field equations, we obtain a family of solutions describing regular black holes spacetimes free from central singularities. Our solutions provide a unified framework that captures both the formation mechanism of black holes through gravitational collapse and their subsequent evolution through Hawking evaporation.

By establishing an exact matching with metrics describing collapsing baryonic matter, we construct a comprehensive physical model that traces the complete life cycle of regular black holes from their formation to their eventual fate. This matching procedure ensures the physical consistency of our solutions while illuminating the detailed dynamics of gravitational collapse. Additionally, we analyze the distinctive shadow characteristics of these regular black holes, deriving specific observational signatures that could potentially distinguish them from their singular counterparts in future astronomical observations.

The paper is organized in a systematic progression through the theoretical framework and its applications. In Section 2, we establish the mathematical foundations of regular black holes, presenting the general formalism that underpins our analysis. This section develops the necessary field equations and derives novel solutions to Einstein's equations that characterize regular black holes with well-defined properties. Section 3 advances the analysis by demonstrating the precise matching conditions between our derived solutions and the Hussain metric, ensuring mathematical consistency and physical relevance.
Section 4 explores the observable implications of these solutions through a detailed investigation of black hole shadow characteristics. The final two sections synthesize our findings: Section 5 presents a comprehensive discussion of the theoretical results and their physical implications, while Section 6 concludes with broader insights into the significance of our work and suggests promising directions for future research.

We use a geometrized system of units in which $c=8\pi G=1$. Also, we adopt the signature $-+++$.

\section{Theoretical framework}

In this section, we briefly examine the conditions required for the formation of a regular black hole. We derive a general formula and identify the criteria under which a regular black hole can emerge from gravitational collapse, considering an arbitrary EoS. To begin, we analyze a metric that represents the most general form of a spherically symmetric dynamical black hole, expressed as:
\begin{equation} \label{eq:metric}
ds^2=-\left(1-\frac{2M(v,r)}{r}\right)dv^2+2\varepsilon dvdr+r^2d\Omega^2.
\end{equation}
Here, \( M(v, r) \) is the mass function, which depends on both the advanced time \( v \) and the radial coordinate \( r \). This function characterizes the dynamical nature of the black hole, allowing for variations in mass due to processes such as accretion or radiation. The parameter \( \varepsilon = \pm 1 \) indicates the direction of the radiation flow, where \( \varepsilon = +1 \) corresponds to ingoing radiation and \( \varepsilon = -1 \) corresponds to outgoing radiation. The coordinate \( v \) represents the advanced time in Eddington-Finkelstein coordinates, which are particularly useful for describing causal structures, such as light rays near the horizon. The angular part of the metric is given by \( d\Omega^2 = d\theta^2 + \sin^2\theta \, d\varphi^2 \), which describes the geometry of a unit two-sphere. This metric setup is essential for studying regular black holes, as it provides a general framework to analyze the radial and time-dependent evolution of the system. Its generality makes it applicable to a wide range of scenarios, including gravitational collapse, black hole evaporation, and the effects of exotic matter fields~\cite{Mkenyeleye:2014dwa, Vertogradov:2018ora, Vertogradov:2018ora, Vertogradov:2020blg, Vertogradov:2022znp, Vertogradov:2023uav, Heydarzade:2023gmd, Vertogradov:2022oif, Herrera:2006ir, Vertogradov:2022yja,Manna:2019gni}. The metric in Eq.~\eqref{eq:metric} is supported by an energy-momentum tensor of the form:
\begin{equation}
T_{ik}=(\rho+P)(l_in_k+l_kn_i)+Pg_{ik}+\mu l_il_k,
\end{equation}
where \( \rho \) and \( P \) denote the energy density and pressure of the matter, respectively. The parameter \( \mu = \varepsilon \frac{2 \dot{M}}{r^2} \) represents the total energy flux. The vectors \( l^i \) and \( n^i \) are null vectors with the following properties:
\begin{eqnarray}
l_{i}&=&\delta^0_i,\nonumber \\
n_{i}&=&\frac{1}{2} \left (1-\frac{2M}{r} \right )\delta^0_{i}-\varepsilon \delta^1_{i},\nonumber \\
l^il_i&=&n^in_i=0,~~ l^in_i=-1.
\end{eqnarray}
The energy density \(\rho\) and pressure \(P\) for this spacetime are given as follows:
\begin{eqnarray} \label{eq:press}
\rho&=&\frac{2M'}{r^2},\nonumber \\
P&=&-\frac{M''}{r}.
\end{eqnarray}
The system of equations \eqref{eq:press} consists of two differential equations and three unknown functions: \(\rho(v, r)\), \(P(v, r)\), and \(M(v, r)\). Therefore, an additional equation must be introduced to close the system, providing three equations for the three unknown functions. \textcolor{black}{ A common way to close the Einstein field equations in a gravitational collapse scenario is to impose an EoS relating the pressure \(P\) to the energy density \(\rho\) as \(P = P(\rho)\). One of the simplest and most widely used choices is the barotropic EoS as \(P = k \rho\), where \(k\) is a constant parameter. In this linear relation, the value of \(k\) determines the effective type of matter: For example:  \(k = 1\): \textit{Stiff fluid}, for which \(P=\rho\). This EoS is often invoked in contexts of extremely high density, such as certain early-universe models or in the study of neutron-star cores. For \(k = \tfrac{1}{3}\): \textit{Radiation}, corresponding to a relativistic gas satisfying \(P = \tfrac{1}{3}\rho\).  This is the appropriate EoS during a radiation-dominated era. For:  \(k = 0\): \textit{Dust}, i.e., pressureless matter with \(P=0\). For \(k = -1\): \textit{Vacuum energy} (cosmological constant), for which \(P = -\rho\) and \(\rho = \text{const}\).  In a strictly dynamical collapse scenario, a true “vacuum fluid” of this form would correspond to adding a fixed cosmological term rather than a collapsing perfect fluid. When one studies self-similar or other classes of collapse solutions with \(P = k\,\rho\), a well-known family of metrics is due to Husain \emph{et al.} (often referred to as “Husain–Martinez–Nuñez” or simply “Husain”) \cite{Husain:1996exact,Husain:1994uj}.  For \(\,0 < k \leq 1\), most Husain-type solutions lead to the formation of a covered (i.e., black-hole) singularity.  However, if \(k\) is sufficiently small (e.g.\ \(0 < k \lesssim 0.01\) in certain self-similar ansätze), it has been shown that a locally naked singularity can form, thereby providing explicit counterexamples to strong cosmic censorship in these models \cite{Joshi:2011rlc,Harada:2001nh}.  Consequently, the end state depends sensitively on \(k\) and on the detailed assumptions of the collapse ansatz. 
In realistic collapse scenarios, the microphysics of the fluid may change as density and temperature evolve.  Thus one often promotes the constant \(k\) to a dynamical field \(k(r,v)\) depending on the comoving radius \(r\) and an advanced (or retarded) time coordinate \(v\).  The generalized barotropic EoS then takes the form
\begin{equation}
P=k(r,v)\rho,
\end{equation}
which accommodates phase transitions (e.g., from radiation-dominated to matter-dominated behavior) or changes in the degrees of freedom (e.g., hadronization of a quark-gluon plasma).  This generalized EoS allows the ratio \(P/\rho\) to vary during different stages of collapse, thereby capturing the evolving properties of the collapsing matter.}

 This generalization allows the system to account for transitions between different phases of matter and provides a more realistic description of gravitational collapse and the potential formation of regular black holes. From Eq.~\eqref{eq:press} and energy-momentum conservation $T^{ik}_{;k}=0$, the following relationship between pressure and energy density can be obtained:
\begin{equation}
P=-\rho-\frac{r}{2} \rho '.
\end{equation}
or
\begin{equation}
\rho 'r=-2P-2\rho
\end{equation}
denotes the radial derivative of the energy density. Substituting this expression into the EoS \(P = k\rho\), one obtains:
\begin{equation}
\rho'r=-(2+2k)\rho.
\end{equation}
The solution of this differential equation is given by:
\begin{equation} \label{eq:sol_den}
\rho=\frac{\rho_0}{r^2}e^{-\int \frac{2k}{r}dr}.
\end{equation}
here \(\rho_0\) is a positive integration constant. This solution represents the energy density \(\rho\) as a function of the radial coordinate \(r\), and its behavior depends on the form of \(k = k(v, r)\).
In general, however, the equation \eqref{eq:sol_den} cannot be integrated analytically. To proceed with the analysis, it is necessary to specify the explicit form of the EoS parameter $k(v,r)$.  To ensure a regular solution at the center (\(r \to 0\)), the parameter \(k\) is expanded as a power series around \(r = 0\):
\begin{equation}
k=\sum_{i=0}^n k_ir^i.
\end{equation}
where the coefficients \(k_i(v)\) are functions of the advanced time \(v\) and are defined as:
\begin{equation}
k_i(v)\equiv \left. \frac{1}{i!}\frac{\partial^i k}{\partial x^i}\right|_{r=0}.
\end{equation}
Then \eqref{eq:sol_den} can be written as
\begin{equation} \label{eq:density1}
\rho=\frac{\rho_0}{r^{2+2k_0}}e^{-2\sum_{i=1}^n k_i\frac{r^i}{i}}.
\end{equation}
To ensure a regular center, it is necessary to demand that \(\rho(0) = \rho_0 = \text{const.}\), meaning the energy density must remain finite at the center. This requirement imposes the following constraint on the parameter \(k_0\):
\begin{equation}
k_0\leq -1.
\end{equation}
If \(k_0 = -1\), the center corresponds to a vacuum medium with a de Sitter core. On the other hand, if \(k_0 < -1\), the energy density vanishes, and the central region transitions to a Minkowski spacetime, indicating an absence of matter. Using Eq.~\eqref{eq:sol_den}, the mass function can be derived in the following general form:
\begin{equation} \label{massgen}
M(v,r)=\frac{\rho_0}{2}\int r^{-2k_0}e^{-2\sum_{i=1}^nk_i\frac{r^i}{i}}+M_0(v).
\end{equation}
where \(M_0(v)\) is an integration function depending on the advanced time \(v\). This function represents the dynamic contribution to the mass function from the evolving system. One critical property of a regular center is that the mass function must vanish at the center, i.e., \(M(v, 0) = 0\). To satisfy this condition and eliminate \(M_0(v)\), it must hold that:
\begin{equation} \label{eq:limit}
\lim\limits_{r\rightarrow 0}\frac{\rho_0}{2}\int r^{-2k_0}e^{-2\sum_{i=1}^n k_i\frac{r^i}{i}}=-M_0(v)
\end{equation}
implying no additional contributions to the mass at the center. This ensures the consistency of the regular center condition and provides a physically meaningful mass function.

\subsection{Model 1: $k_0=-1, k_3=k_3(v)$}

\begin{figure}[h!]
    \centering
    \includegraphics[width=0.8\linewidth]{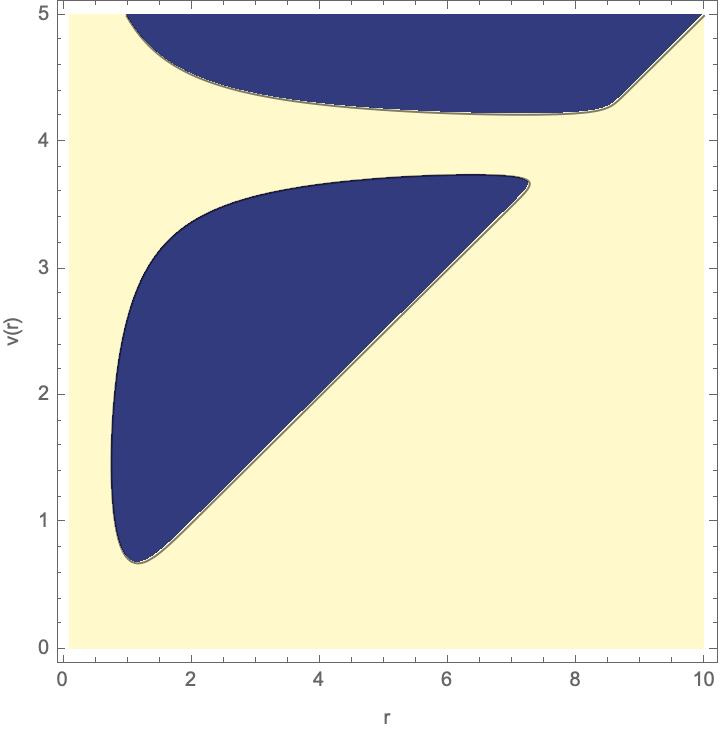}
    \caption{\textcolor{black}{Setting $F(v,r)=0$ with $\mu=4$ and $\lambda=1$ defines the loci of the inner and outer apparent horizons, as shown in the above figure. This configuration models a black hole that forms and then evaporates. Initially, the outer apparent horizon is spacelike and grows in radius, while the inner apparent horizon is timelike and shrinks. At the radial location where the energy conditions are first violated, the outer horizon attains its maximum extent and the inner horizon reaches its minimum. Beyond this boundary, the outer horizon becomes timelike and begins to shrink, whereas the inner horizon becomes spacelike and starts to expand. At a critical value of $v$, the two horizons coincide forming an extremal configuration and then vanish entirely, leaving a horizonless, regular central region. Shortly thereafter, both horizons reappear in a regime where the energy conditions are everywhere satisfied. In this later phase, matter accretion dominates: the outer horizon again expands while the inner horizon contracts, signifying continued growth of the black hole. }}
    \label{fig1}
\end{figure}

\begin{figure}[h!]
    \centering
    \includegraphics[width=0.8\linewidth]{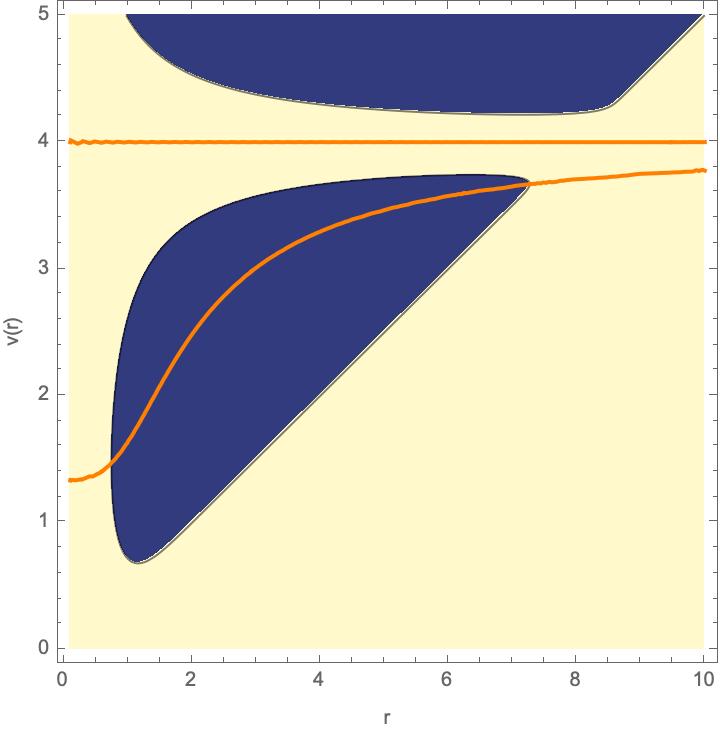}
    \caption{\textcolor{black}{Same as Figure~\ref{fig1}, but with \(\dot{M} = \frac{dM}{dv} = 0\) shown by the orange curve. In this case, the minimum of the outer apparent horizon and the maximum of the inner apparent horizon coincide exactly with the NEC horizon.}
}
    \label{fig2}
\end{figure}

The integral in the formula \eqref{massgen} can be simplified only for a specific set of parameters \(k_i\). In most cases, the integration leads to expressions involving gamma functions or other special functions, making explicit analysis challenging. As a result, we were able to identify only two particular solutions that can be integrated in closed form without invoking gamma functions, which are presented in this work. Among these cases, we are particularly interested in regular solutions with a de Sitter core at the center. Such solutions correspond to cases where the parameter choices allow for a finite energy density and a well-defined spacetime structure at the core. Notably, only a few parameter configurations result in a tractable form for the mass function. In this study, we focus on the simplest case where \(k_1 = k_2 = 0\). This choice eliminates the higher-order terms in the expansion of \(k(v, r)\), significantly simplifying the calculation. By applying the method described in the previous section, the mass function can be expressed as:
\begin{equation} \label{eq:mass}
    M(v, r) = \frac{\rho_0(v)}{2} \int e^{-2 \int \frac{k(v, r)}{r} \, dr} \, dr + M_0(v),
\end{equation}
where \(\rho_0(v)\) is the central energy density as a function of advanced time \(v\), and \(M_0(v)\) is an additional integration function of \(v\), representing the overall black hole mass. The term \(M_0(v)\) is typically determined by boundary conditions or asymptotic properties of the black hole. This formulation emphasizes the role of \(k(v, r)\) in defining the structure of the mass function and ensures that the solution remains regular at the center. The case \(k_1 = k_2 = 0\) is particularly relevant as it corresponds to a de Sitter core, which is a common feature in regular black hole models, providing a smooth and non-singular interior spacetime. After calculating the Kretschmann scalar for the generalized Vaidya spacetime, it is expressed as:
\begin{eqnarray} \label{eq:scalar}
       K = \frac{4}{r^6} [ 4 \left( 3M^2 - 4rMM' + 2r^2M'^2 \right) \\ + 4r^2M''\left(M - rM'\right)  + 4r^4M''^2 ] \notag,
\end{eqnarray}
where \(M = M(v, r)\) is the mass function, \(M' = \frac{\partial M}{\partial r}\), and \(M'' = \frac{\partial^2 M}{\partial r^2}\). To ensure regularity at the center (\(r \to 0\)), we require that the Kretschmann scalar remains finite:
\begin{equation}
    \lim\limits_{r \to 0} K \neq \infty.
\end{equation}
This condition ensures that there are no curvature singularities at the center. Substituting the series expansion for \(k(v, r)\)) into the mass function \(M(v, r)\) (from Eq.~\eqref{eq:mass}) and the Kretschmann scalar \(K\), we find that the coefficients \(k_i(v)\) must satisfy certain constraints to ensure the finiteness of \(K\) as \(r \to 0\). Specifically, the following condition must hold:
\begin{equation} \label{eq:condition}
    M_0(v) = -\lim\limits_{r \to 0} \frac{\rho_0(v)}{2} \int e^{-2 \int \frac{k(v, r)}{r} \, dr} \, dr,
\end{equation}
where \(M_0(v)\) is the integration function related to the black hole mass. This constraint ensures that the central mass contribution is well-defined and eliminates divergences in the Kretschmann scalar at the center. The condition ties the behavior of the coefficients \(k_i(v)\) in the expansion of \(k(v, r)\) to the structure of the spacetime, providing a consistent framework for describing regular black holes. We begin by assuming the following specific parameter choices:
\begin{eqnarray} \label{eq:alpha1}
k_0(v) &=& -1, \nonumber \\
k_3(v) &=& k_3(v), \nonumber \\
k_1(v) &=& k_2(v) = 0.
\end{eqnarray}
With these assumptions, the mass function takes the form:
\begin{equation} \label{eq:mass1}
M(v, r) = M_0(v) - \frac{3\rho_0(v)}{2k_3(v)} e^{-\frac{1}{9}k_3(v)r^3}.
\end{equation}
Now we consider different cases of \( M_0(v)\): 
\begin{itemize}
    \item \textbf{Case 1: \( M_0(v) \equiv \frac{3\rho_0(v)}{2k_3(v)} \):}
\end{itemize}
    In this case, the spacetime described by the metric \eqref{eq:metric} with the mass function \eqref{eq:mass1} represents a regular black hole, which is a dynamical generalization of the well-known Dymnikova regular black hole~\cite{bib:dym1}. The Kretschmann scalar for this spacetime at \(r = 0\) is given by:
\begin{equation}
\lim\limits_{r \to 0} K = \frac{32}{27} M_0(v)^2 k_3(v)^2.
\end{equation} This ensures that the spacetime curvature at the center remains finite, consistent with the requirements of a regular black hole. For this model, the energy density \(\rho\) and pressure \(P\) are expressed as:
\begin{eqnarray} \label{eq:density_dym}
    \rho &=& \frac{2}{3} M_0(v) k_3(v) e^{-\frac{1}{9}k_3(v)r^3}, \nonumber \\
    P &=& \frac{2}{3} M_0(v) k_3(v) e^{-\frac{1}{9}k_3(v)r^3} \left( \frac{k_3(v) r^3}{6} - 1 \right).
    \end{eqnarray}
These expressions reveal the distribution of matter and its dynamics in the spacetime. The exponential term ensures that both the energy density and pressure decay smoothly away from the center, contributing to the regularity of the black hole. The weak and dominant energy conditions impose the following constraints on the parameters of the solution:
\begin{equation}
    \rho_0(v) > 0,
\end{equation}
\begin{equation}
    r \leq \left( \frac{12}{k_3} \right)^{\frac{1}{3}}.
\end{equation}
This solution exhibits a \textit{de Sitter core} at the center (\(r \to 0\)) and approaches the \textit{Schwarzschild limit} at infinity (\(r \to \infty\)). For specific choices of the functions:
\begin{eqnarray}
k_3(v) &=& (\mu - v)^2, \nonumber \\
\rho_0(v) &=& \frac{2}{3} \lambda k_3(v) v,
\end{eqnarray}
we can visualize the behavior of the spacetime by plotting the function \(F(v, r) = 1 - \frac{2M(v, r)}{r} = 0\), which represents the horizon structure. In Figure~\ref{fig1}, \(F(v, r) = 0\) is plotted as \(v(r)\), with parameters \(\mu = 4\) and \(\lambda = 1\). Then we also plot the graph of \(\dot{M} = 0\) in Figure~\ref{fig2}.

\begin{itemize}
 \item \textbf{Case 2: \(M_0(v) \neq \frac{3\rho_0(v)}{2k_3(v)}\):}
\end{itemize}
In this case, the spacetime described by the metric \eqref{eq:metric} with the mass function \eqref{eq:mass1} corresponds to a \textit{singular black hole}. The Kretschmann scalar diverges as \(r \to 0\), indicating the presence of a curvature singularity at the center. It is possible that during the evolution, the function \(N(v) \equiv \frac{c_1(v)}{3f(v)} - c_2(v)\) becomes zero only at specific values of \(v\), such as \(v_1, v_2, v_3, \ldots\). This implies that the system oscillates between states with \emph{regular} and \emph{singular} central regions at discrete instants. Such behavior was first reported in \cite{bib:ali2024CQGnew}. At first glance, this behavior at the center appears absurd, as the singularity alternately forms and disappears. While the disappearance of the singularity is undoubtedly linked to the violation of null energy conditions, such violations typically occur during processes like the accretion of charged particles or Hawking radiation, which do not explain the observed oscillatory behavior of the singularity. The resolution to this paradox lies in the fact that the singularity in this model is a gravitationally weak. This means that infalling objects do not experience its presence and are not disrupted until they reach the exact center. Consequently, the presence or absence of a gravitationally weak singularity only becomes significant at the very center, with no observable effects in its immediate vicinity. The situation would be drastically different if the singularity were gravitationally strong. In that case, any object approaching it would be torn apart into a thin string due to extreme tidal forces. If such a singularity were to oscillate, it would lead to absurd consequences: at one moment, objects would be destroyed, and at another, they would remain intact. However, as noted earlier, in the case of a gravitationally weak singularity, these oscillations have no measurable impact on infalling objects, which is precisely the scenario we are considering here. In Figure~\ref{fig1}, the apparent horizon behavior is illustrated. The first model (Eq.~\eqref{eq:mass1}) violates the null energy condition (NEC) near the center. As shown in \cite{bib:vertogradov2024horizon}, the \textit{outer apparent horizon} is spacelike, while the \textit{inner apparent horizon} is timelike if the NEC horizon is located within the inner horizon. The \textit{outer horizon} grows, while the \textit{inner horizon} shrinks. At the minimum of the inner horizon, it meets the NEC horizon and becomes spacelike, resulting in two growing horizons. At the maximum of the outer horizon, it becomes null, meets the NEC horizon, and transitions to being timelike. Eventually, the two horizons merge and disappear, leaving behind a \textit{horizonless object with a regular center} for a brief period. In Figure~\ref{fig2}, the evolution of NEC horizons is depicted. The NEC horizon meets the inner horizon at its minimum and the outer horizon at its maximum.

\subsection{Model 2: \( k_0 = -1,~~k_1 = k_1(v) \)}

In this section, we consider a model where the parameters satisfy:
\begin{eqnarray}
    k_0(v) &=& -1, \nonumber \\
    k_1(v) &=& k_1(v), \nonumber \\
    k_3(v) &=& k_2(v) = 0.
\end{eqnarray}
Under these conditions, the mass function takes the form:
\begin{equation}
\label{eq:mass2}
\begin{split}
    M(v,r) &= -\frac{\rho_0(v)}{8k_1^3(v)} e^{-2k_1(v)r} [ 1 + 2rk_1(v) + 2r^2k_1^2(v) ] \\
    &\quad + M_0(v),
\end{split}
\end{equation}
where \( k_1(v) \) is an arbitrary function of time, and it must satisfy \( k_1(v) \neq 0 \). This solution generalizes the black hole solution obtained in~\cite{bib:ali2024CQGnew}. Similar to the previous model, we analyze different cases of this spacetime.

\subsubsection{Regular Black Hole Case: \( M_0(v) \equiv \frac{\rho_0(v)}{8k_1^3(v)} \)}

\begin{figure}[h!]
    \centering
\includegraphics[width=0.8\linewidth]{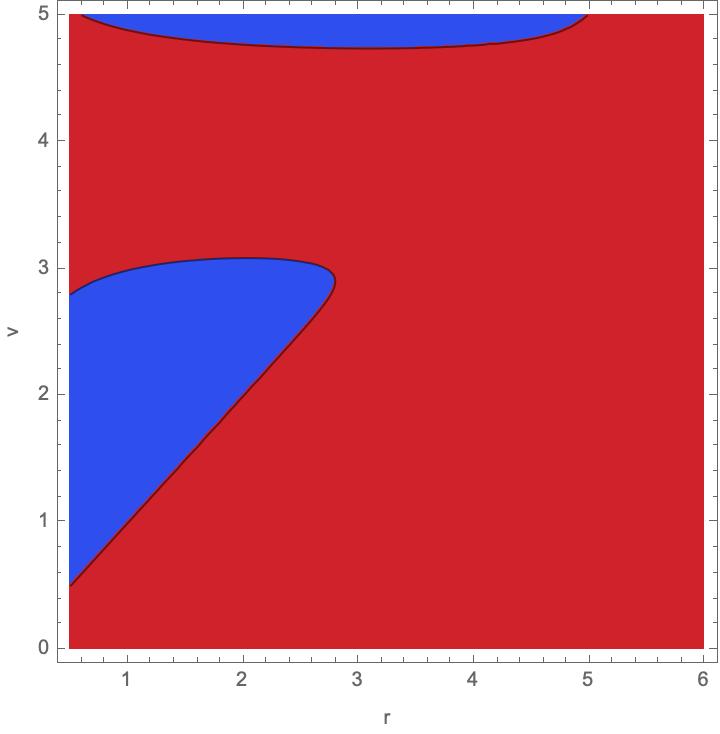}
    \caption{$F(v,r)=0$ for $\mu=4$, $\lambda=1$. The behavior of apparent horizons conside with apparent horizons  in previous model (see \ref{fig1} for detailed discussion).}
    \label{fig3}
\end{figure}

\begin{figure}[h!]
    \centering
\includegraphics[width=0.8\linewidth]{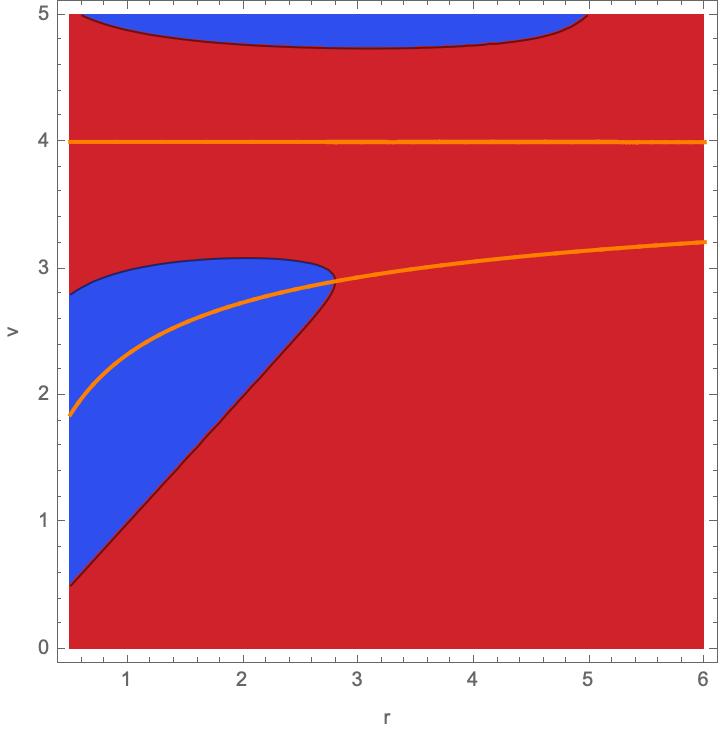}
    \caption{Same graph as Figure~\ref{fig3}, but with $\dot{M} = 0$.}
    \label{fig4}
\end{figure}

In this case, the metric \eqref{eq:metric} with the mass function \eqref{eq:mass2} describes a \textit{dynamical regular black hole}. The Kretschmann scalar at the center \( r = 0 \) is given by:
\begin{equation}
    \lim\limits_{r \to 0} K = \frac{512}{3} k_1^6(v) M_0^2(v).
\end{equation}
This ensures that the spacetime curvature remains finite at the core, satisfying the regularity condition. The energy density \( \rho \) and pressure \( P \) take the following forms:
\begin{eqnarray} \label{eq:density_ali}
    \rho &=& 8M_0(v)k_1^3(v) e^{-2k_1(v)r},\nonumber \\
    P &=& 8M_0(v) k_1^3(v) e^{-2k_1(v)r} (-1 + k_1(v)r).
\end{eqnarray}
These expressions demonstrate that the energy density and pressure decrease smoothly away from the center, contributing to the regular structure of the black hole. The weak and dominant energy conditions require the following constraints:
\begin{equation}
    \rho_0(v) > 0,
\end{equation}
\begin{equation}
    r \leq \frac{2}{k_1(v)}.
\end{equation}
These conditions ensure that the matter distribution satisfies physically reasonable energy bounds. At \( r \to 0 \), the solution exhibits a \textit{de Sitter core}, preventing the formation of a singularity. At \( r \to \infty \), the solution smoothly transitions to the \textit{Schwarzschild limit}, ensuring the expected asymptotic flatness. This model presents an alternative scenario for regular black holes, where the choice of \( k_1(v) \) determines the dynamical evolution of the solution. Under the following choices for the arbitrary functions:
\begin{eqnarray}
    k_1(v) &=& (\mu - v)^2, \nonumber \\
    \rho_1(v) &=& 4\lambda k_1^3(v)v,
\end{eqnarray}
we can, similar to Model 1, plot the function:
\begin{eqnarray}
F(v, r) = 1 - \frac{2M(v, r)}{r} = 0.
\end{eqnarray}
In Figure~\ref{fig3}, the apparent horizon behavior is illustrated. In Figure~\ref{fig4}, the evolution of NEC horizons is depicted. 

\textcolor{black}{We have considered two models of gravitational collapse leading to the formation of a regular black hole. It is important to note that other combinations of the coefficients $ K_i(v) $ can also be considered. However, further analysis would be significantly complicated due to the appearance of gamma functions upon integrating the Einstein equations. The two models presented here are the simplest from an analytical perspective.}

\section{Matching the Interior and Exterior Solutions}

While the solutions \eqref{eq:mass1} and \eqref{eq:mass2} describe a regular center, they are generally valid only in the vicinity of the center. Even though these solutions correspond to known models \cite{bib:dym1, bib:ali2024CQGnew}, a more realistic approach would be to match these solutions to an exterior spacetime representing collapsing matter with a physically realistic EoS. The simplest dynamical exterior solution is the Vaidya metric, but matching in this case is only possible if the energy density in \eqref{eq:mass1} and \eqref{eq:mass2} vanishes at some radius, which is not observed in our solutions. Instead, we consider a \textit{generalized Vaidya metric}, namely the Husain solution, which corresponds to the barotropic EoS:
\begin{eqnarray}
P = \alpha \rho.
\end{eqnarray}
The mass function is given by~\cite{Husain:1996exact, Vertogradov:2016gqc}:
\begin{equation} \label{eq:massh}
    M(v,r) = M_1(v) + \rho_1 r^{1 - 2\alpha}.
\end{equation}
\textcolor{black}{The curvature invariants Ricci scalar (\( R \))}, squared Ricci scalar (\( S \)), and Kretschmann scalar (\( K \)) can be expressed in terms of the mass function \( M(v, r) \), energy density \( \rho(v, r) \), and pressure \( P(v, r) \) as:
\begin{eqnarray} \label{eq:carvature2}
    R &=& 2\rho - 2P, \nonumber \\
    S &=& 2\rho^2 + 2P^2,  \\
    K &=& \frac{48M^2}{r^6} - \frac{16M}{r^3} \left(2\rho - P\right) + 8\rho^2 - 8\rho P + 4P^2.\nonumber
\end{eqnarray}
These expressions indicate that for a smooth matching of all curvature invariants, it is sufficient to match the mass function \( M(v, r) \), the energy density \( \rho(v, r) \), and the pressure \( P(v, r) \) at the matching hypersurface $\sigma$.

\subsection{Dymnikova Solution}

Let us first consider the solution \eqref{eq:mass1}, which corresponds to the \textit{dynamical Dymnikova black hole solution}. Since we are focusing on regular black hole solutions, we must ensure the matching conditions hold. From the expressions for energy density and pressure \eqref{eq:density_dym}, we observe that, in terms of \( k_0 \) and \( k_3 \), the EoS takes the form:
\begin{equation}
    P = \rho \left( \frac{M_0 k_3}{6} - 1 \right).
\end{equation}
This EoS behaves \textit{like a barotropic EoS} \( P = \alpha \rho \) at a specific matching radius \( r = r_b \), where:
\begin{equation} \label{eq:rb_dym}
    r_b^3 = \frac{6(\alpha + 1)}{k_3}.
\end{equation}
We will use this matching radius \( r_b \) to match the \textit{interior Dymnikova solution} with the \textit{exterior Husain solution}. In the Husain solution, the energy density is given by:
\begin{equation} \label{eq:den_hus}
    \rho_{\text{Husain}} = 2(1 - 2\alpha) \frac{\rho_1(v)}{r^{2\alpha + 2}}.
\end{equation}
To ensure a smooth transition between the two solutions at the hypersurface \( r = r_b \), we must match the energy densities and pressures at this radius. This approach provides a physically consistent model where the regular interior solution transitions smoothly into an exterior generalized Vaidya spacetime, allowing for a realistic description of a collapsing regular black hole. The energy density from the Husain solution \eqref{eq:den_hus} matches the energy density of the Dymnikova-like solution \eqref{eq:density_dym} if:
\begin{equation} \label{eq:defd1}
    \rho_1(v) = \frac{M_0(v)}{3(1 - 2\alpha)} r_b^{2\alpha + 2} e^{-\frac{2(\alpha+1)}{3}}.
\end{equation}
Similarly, the mass function in \eqref{eq:mass1} transitions smoothly to the Husain mass function \eqref{eq:massh} at \( r = r_b \) if:
\begin{equation} \label{eq:defc1}
    M_1(v) \big|_{r=r_b} = M_0(v) \left( 1 - \frac{3}{1 - 2\alpha} e^{-\frac{2(\alpha+1)}{3}} \right).
\end{equation}
Thus, the metric tensor, energy density, and pressure of solutions \eqref{eq:mass1} and \eqref{eq:massh} are equal at \( r = r_b \) if conditions \eqref{eq:defd1} and \eqref{eq:defc1} hold. This confirms that the Husain solution can describe the gravitational collapse leading to a regular black hole formation.

\subsection{Black hole with Hagedorn fluid}

We now apply the same method to match the regular black hole solution \eqref{eq:mass2} with the Husain solution \eqref{eq:massh}. From the energy density \eqref{eq:density_ali}, the EoS follows:
\begin{equation}
    P = (k_1 r - 1) \rho.
\end{equation}
This behaves like a barotropic EoS \( P = \alpha \rho \) at the matching radius \( r = r_b \), where:
\begin{equation} \label{eq:rb_ali}
    r_b = \frac{\alpha + 1}{k_1}.
\end{equation}
At this radius, the energy density \eqref{eq:density_ali} transforms into the Husain energy density \eqref{eq:den_hus} if:
\begin{equation} \label{eq:defd2}
    \rho_1 = \frac{4M_0 k_1^3}{1 - 2\alpha} r_b^{2\alpha + 2} e^{-2(\alpha+1)}.
\end{equation}
Additionally, the \textit{mass function} in \eqref{eq:mass2} transitions to the Husain solution \eqref{eq:massh} at \( r = r_b \) if the condition:
\begin{equation} \label{eq:defc2}
    M_1 = M_0 \left( 1 - \frac{9 + 6\alpha + 5\alpha^2 + 2\alpha^3}{1 - 2\alpha} e^{-2(\alpha+1)} \right)
\end{equation}
is satisfied. Ensuring these conditions, the dynamical collapse of the Husain solution leads to the formation of a regular black hole with a smooth transition from the interior to the exterior spacetime. The solution \eqref{eq:mass2} is successfully matched with the \textit{Husain solution} \eqref{eq:massh} at the hypersurface \( r = r_b \) (as defined in \eqref{eq:rb_ali}) if the conditions \eqref{eq:defd2} and \eqref{eq:defc2} hold. It is important to note that, in this case, matching solution \eqref{eq:mass2} with the standard Vaidya spacetime is not possible because, for any \( r > 0 \), the energy density of the Husain solution remains nonzero:
\begin{equation}
    \rho_{\text{Husain}} \neq 0.
\end{equation}
If one opts to use the Vaidya spacetime as an exterior solution for these metrics, it would be necessary to introduce a thin matter layer at the matching hypersurface. In this case: The interior solutions \eqref{eq:mass1} or \eqref{eq:mass2} must be matched with a Vaidya spacetime. A thin matter layer would be required at the matching interface to account for the discontinuity in the energy density. However, performing such a matching lies beyond the scope of this paper, so we do not explore this possibility further in the current study.

\section{Shadow properties}

\textcolor{black}{When examining observational signatures of our collapse model, one should analyze the shadow properties of the exterior Husain solution. In principle, the dynamical shadow of the Husain metric (Eq.~\eqref{eq:massh}) could be computed using the method developed in Ref.~\cite{Vertogradovalidynamical}. However, that approach is valid only for slowly evolving spacetimes and cannot be applied during rapid gravitational collapse. For this reason, we instead calculate the black‐hole shadow of the \emph{static} Husain solution, which describes the fully formed black hole. This metric can be written in the form:}

\begin{eqnarray} \label{hus}
ds^2&=&-\left(1-\frac{2M}{r}+\frac{J}{r^{2\alpha}}\right)dt^2+\left(1-\frac{2M}{r}+\frac{J}{r^{2\alpha}}\right)^{-1}dr^2 \notag \\&+&r^2d\Omega^2.
\end{eqnarray}
Here, \(M\) denotes the black hole mass, and \(J\) is a parameter encoding the properties of the baryonic matter (in the special case \(\alpha = 1\), one has \(J = Q^2\), so that the metric reduces to the Reissner–Nordström solution). The constant \(\alpha\) specifies the barotropic equation of state \(P = \alpha \rho\). To ensure that the weak, strong, and dominant energy conditions are all satisfied throughout the spacetime, the following inequalities must hold:
\begin{eqnarray}
\alpha \in [-1, 1],~~ \alpha \neq \frac{1}{2},\nonumber \\
J\leq M,~~ J>0 ~~\text{if}~~ \alpha >\frac{1}{2}.
\end{eqnarray}
The shadow radius is calculated as follows \cite{Vagnozzi:2022moj}
\begin{eqnarray}
R_{sh}&=&\frac{r_{ph}}{\sqrt{f(r_{ph})}},
\end{eqnarray}
where the photon sphere radius is calculated by solving this relation for $r_{ph}$:   \begin{eqnarray}f'(r_{ph})r_{ph}&=&2f(r_{ph}).\end{eqnarray}

\textcolor{black}{It should be noted that, in our analysis, both the photon sphere radius and the black hole shadow are determined by considering only null geodesics confined to the equatorial plane, thereby neglecting any off plane light deflection.
}

\begin{table}[h!]
\centering
\begin{tabular}{|c|c|c|}
\hline
$\alpha$ & $r_{\text{ph}}$ & \textcolor{black}{$R_{\text{sh}}$} \\ \hline
1 & 2.61803 & 4.70960 \\ \hline
2 & 2.94104 & 5.14587 \\ \hline
3 & 2.99165 & 5.19076 \\ \hline
4 & 2.99885 & 5.19556 \\ \hline
5 & 2.99985 & 5.19609 \\ \hline
6 & 2.99998 & 5.19615 \\ \hline
\end{tabular}
\caption{\textcolor{black}{Photon sphere radius ($r_{\text{ph}}$) and shadow radius ($r_{\text{sh}}$) for different values of $\alpha$ with constant $M=1$ and $J=0.5$.}}
\label{tab:shadow}
\end{table}

\begin{figure}[h!]
    \centering
\includegraphics[width=1\linewidth]{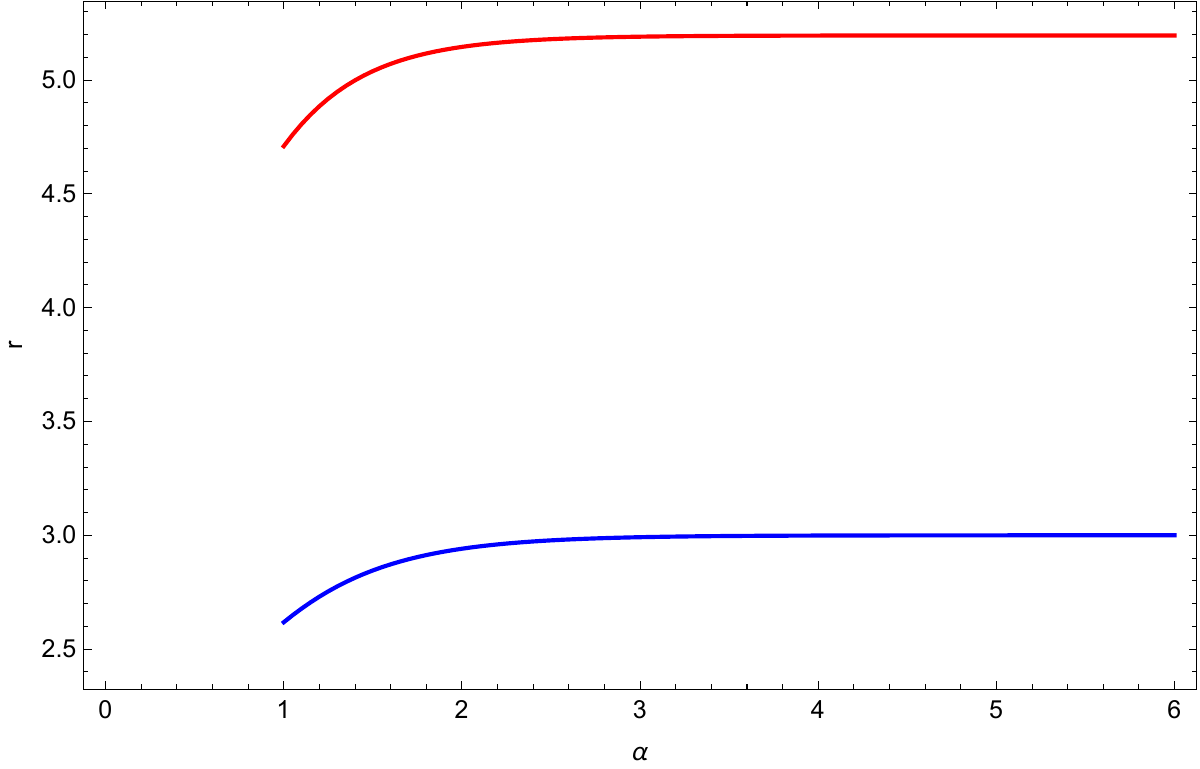}
    \caption{Figure shows the shadow radius and photon sphere radius versus $\alpha$.}
    \label{shadowfig}
\end{figure}

\begin{figure}[h!]
    \centering
\includegraphics[width=1\linewidth]{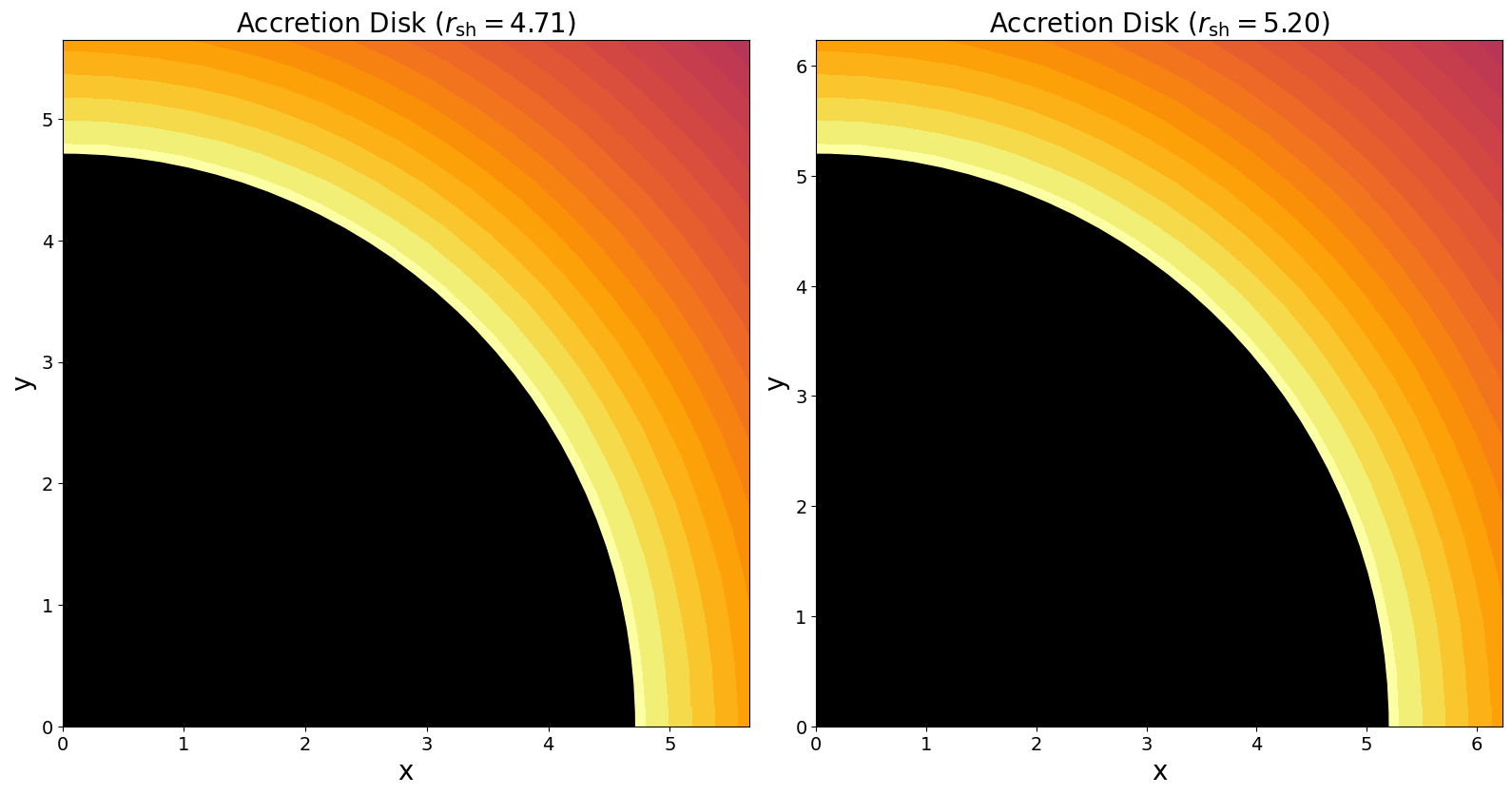}
    \caption{\textcolor{black}{Figure shows the spherical accetion disk around black hole for different shadow radiuses.}}
    \label{shadowfig2}
\end{figure}

\textcolor{black}{From Table~\ref{tab:shadow} and Figures \ref{shadowfig}, \ref{shadowfig2}, we observe that as the parameter $\alpha$ increases, both the photon sphere radius $r_{\text{ph}}$ and the shadow radius $R_{\text{sh}}$ increase monotonically.  As $\alpha$ grows, the most substantial changes occur for moderate values 1 $\le \alpha \le 3$, after which both radii begin to saturate around $r_{\text{ph}}\approx 3.0$ and $R_{\text{sh}}\approx 5.20$. Physically, a larger photon sphere and shadow radius imply enhanced light bending, allowing photons to orbit at larger radii and causing a more extended dark region, or ``silhouette,’’ as seen by a distant observer. The near-constant values for $\alpha \ge 4$ suggest an asymptotic regime in which further increases in $\alpha$ have diminishing effects on the geometry, pointing to a limiting configuration for the underlying spacetime model.}

\textcolor{black}{In our collapse model, the non-singular interior solution is only realized within a finite central region of the spacetime after the fluid has undergone the phase transition.  Outside this region, one must match continuously onto the Husain metric to obtain a globally well-defined, dynamical black hole.  Since the photon sphere and shadow observables are determined by null geodesics that travel through regions well outside the central core, they are effectively governed by the exterior Husain geometry in Eq. \ref{hus}.  Consequently, for the purpose of computing the shadow radius and photon sphere radius, we employ only the Husain solution. }

\section{Discussions}

\begin{figure}[h!] 
    \centering
    \includegraphics[width=0.6\linewidth]{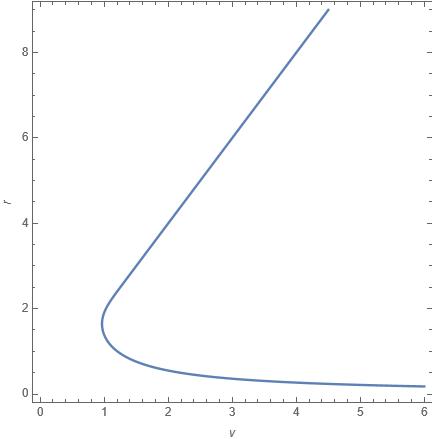}
\caption{\textcolor{black}{The plot demonstrates the behavior of apparent horizons under the validity of energy conditions. Here, $M_0(v) = K_3(v) = v$. In this case, the energy conditions are satisfied throughout the entire spacetime: the outer horizon is a spacelike hypersurface and expanding, while the inner horizon is a timelike hypersurface and contracting.}}
    \label{fig5}
\end{figure}

\begin{figure}[h!]
    \centering    \includegraphics[width=0.6\linewidth]{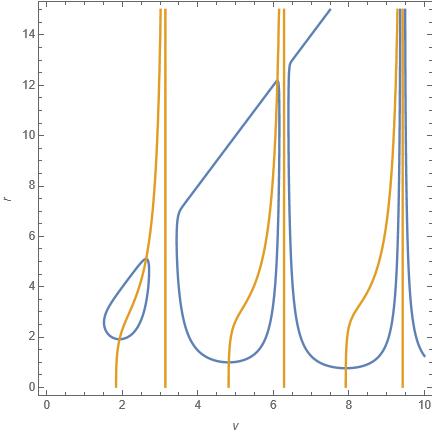}
\caption{
\textcolor{black}{This plot illustrates the behavior of horizons for the mass function given by \eqref{Eq52}, where $ M_0(v) = v $ and $ K_3(v) = \sin^2 v $. Since the function $ K_3(v) $ vanishes at multiple points, the energy conditions are violated near the center, and this region expands over time.}}
 \label{fig6}
\end{figure}

In this section, we analyze the obtained solutions and the constraints they must satisfy. Our discussion primarily focuses on the solution:
\begin{equation}
    M(v,r) = M_0(v) \left( 1 - e^{-\frac{1}{9}k_3(v)r^3} \right),
\end{equation}
though the conclusions drawn here also apply to the second model. One of the key observations is that when the function \( k_3(v) \) increases, the energy conditions remain valid, and the black hole possesses two apparent horizons: Outer apparent horizon: Spacelike and increasing. Inner apparent horizon: Timelike and shrinking. This behavior follows from the condition:
\begin{equation}
    \dot{M} = \dot{M}_0 - \dot{M}_0 e^{-k_3 r^3} + \dot{k}_3 M_0 r^3 e^{-k_3 r^3} \geq 0.
\end{equation}
However, the dynamics change dramatically when \( k_3(v) \) starts decreasing. In this case: The energy conditions are violated, and the structure of the apparent horizons changes when the NEC horizon crosses one of the apparent horizons. The region of energy condition violation expands and extends to infinity as \( k_3(v) \to 0 \). At this point, the apparent horizons merge and disappear. However, as shown in the Figures~\ref{fig5} and~\ref{fig6} they reappear later, with the outer apparent horizon growing and the inner horizon shrinking. This happens because when the horizons reappear, the energy conditions are no longer violated. As an example, we consider a mass function of the form:
\begin{equation} \label{Eq52}
    M(v,r) = M_0(v) - M_0(v) e^{-r^3 \sin^2 v}.
\end{equation}
This type of function illustrates the periodic nature of horizon formation and disappearance. 

\textcolor{black}{The evolution of the apparent horizons is illustrated in Figure~\ref{fig5} for the regime in which the energy conditions remain satisfied. Initially, the inner apparent horizon shrinks until it reaches a minimum and becomes null; it then transitions to a spacelike hypersurface and begins to expand. This change signals that the inner horizon has entered a region where the energy conditions are violated. Meanwhile, the outer horizon remains spacelike as it grows. Once the outer horizon attains its maximum and becomes null, it transitions to a timelike hypersurface and begins to contract. This contraction indicates that the outer horizon now lies in a region of violated energy conditions, signaling the onset of black hole evaporation. As a consequence, the two horizons approach one another and eventually merge, forming an extremal configuration before disappearing altogether. The subsequent plots in Figure~\ref{fig6} display qualitatively similar dynamics. We emphasize that one should focus on horizon behavior only up to the first horizontal line (i.e., until the first zero of the sine function in \(k_{3}(v)\)). Beyond that point, the plots reveal an infinite sequence of apparent horizon formations and disappearances-each time \(k_{3}(v)\) vanishes. The vertical lines in Figures~\ref{fig5} and \ref{fig6} mark these zeros of \(k_{3}(v)\), where the black hole effectively “disappears.” However, there is no physically plausible mechanism that could drive a continuous cycle of black hole formation and disappearance. Specifically, when $k_{3}(v) \;=\; 0\,,$ the spacetime undergoes a discontinuous transition: the region that previously violated the energy conditions suddenly becomes one in which all energy conditions hold. Such a jump-of the second kind-is unphysical. Therefore, the evolution must be truncated at the first moment when \(k_{3}(v) = 0\). Any further continuation beyond this point would lead to unexplainable and unphysical behavior. This requirement imposes a fundamental constraint on our model, ensuring that the collapse remains physically meaningful. Moreover, recall that the interior solutions given by Eqs.~\eqref{eq:mass1} and \eqref{eq:mass2} were obtained via a near-center expansion of \(k(v,r)\). As a result, they are only valid up to the radius where the pressure begins to increase with \(r\), which inevitably leads to a violation of the dominant energy condition (DEC). This violation occurs at a critical radius \(r_{\rm violation}\), beyond which the model can not be trusted. In a realistic gravitational collapse of baryonic matter, the energy density and temperature rise sharply; at sufficiently high temperatures (e.g.\ \(T \sim 10^{15}\,\mathrm{K}\), the grand unification scale), the description in terms of ordinary baryonic matter breaks down. One must then consider the collapse of a different, high-energy matter phase. The transition radius \(r_{b}\) at which the phase change occurs depends both on the initial baryonic properties (parametrized by \(\alpha\)) and on the parameters \(k_{3}\) or \(k_{1}\) that describe the post-transition phase in models \eqref{eq:mass1} and \eqref{eq:mass2}, respectively. For massive stars collapsing without a supernova explosion, \(r_{b}\) can be as large as \(\sim 1\,\mathrm{km}\) when \(\alpha = 1\) (the Bonnor-Vaidya limit). As a concrete example, consider the star R136a1 in the Tarantula Nebula of the Large Magellanic Cloud, with
$M_{\mathrm{R136a1}} \approx 315\,M_{\odot}$ and 
$R_{\mathrm{R136a1}} \approx 35\,R_{\odot}.$
These estimates justify the necessity of the matching procedure since at radii on the order of kilometers (well outside the near-center expansion’s domain of validity), the matter has already undergone a phase transition and the DEC is violated. Accordingly, one must join the interior solution to an appropriate exterior geometry at \(r = r_{b}\) to construct a globally consistent collapse model.}

\section{Conclusions}

\textcolor{black}{The solutions given by Eqs.~\eqref{eq:mass1} and \eqref{eq:mass2} yield a regular black hole, thereby illustrating that singularity formation can be avoided. However, they exhibit a crucial drawback: the pressure grows with increasing radial coordinate \(r\). This anomalous pressure profile gives rise to two major problems. First, it is physically implausible for the pressure to increase toward the outer layers of a collapsing star. Second, at a critical radius \(r = r_{\text{violation}}\), the Dominant Energy Condition (DEC) is violated. The DEC violation indicates that these interior solutions must be matched to an appropriate exterior spacetime beyond \(r_{\text{violation}}\) to maintain a physically consistent global geometry.}

\textcolor{black}{Furthermore, the interior solutions given by Eqs.~\eqref{eq:mass1} and \eqref{eq:mass2} were derived via a power-series expansion of the EoS parameter \(k(v,r)\) around \(r=0\). As a consequence, these expressions are valid only in the immediate vicinity of the regular center and cannot be extended to the entire spacetime. To obtain a complete, globally well-defined dynamical black hole model, one must therefore match the interior solution to a suitable exterior geometry. The simplest candidate for a radiating, spherically symmetric exterior is the Vaidya metric. However, demanding a smooth junction between Eq.~\eqref{eq:mass1} and Eq.\eqref{eq:mass2} and the Vaidya spacetime generally requires inserting a thin shell of additional matter at the matching surface. This extra layer complicates the construction and makes the overall model appear more contrived. }

\textcolor{black}{A more natural physical interpretation is that the solutions in Eqs.~\eqref{eq:mass1} and \eqref{eq:mass2} describe the response of ordinary baryonic matter under critical compression during collapse. In this view, the exterior region must itself satisfy a barotropic EoS \(P = \alpha \rho\), which is precisely the Husain solution \cite{Husain:1996exact}. By imposing standard junction conditions at some matching radius \(r = r_{\rm match}\), we smoothly join our near-center expansions Eq. \eqref{eq:mass1} and Eq. \eqref{eq:mass2} to the Husain metric. This construction yields a single, globally well-defined dynamical black hole spacetime in which the collapse of baryonic matter naturally produces a regular central core.}

\textcolor{black}{To explore potential observational signatures, we computed the black hole shadow, restricting attention to the dynamical Husain solution rather than its static limit. Nevertheless, it remains an open question whether every collapse leads to a regular center or if a singularity can still form. From Table~\ref{tab:shadow} and Figures~\ref{shadowfig} and \ref{shadowfig2}, we find that, as the EoS parameter \(\alpha\) increases, both the photon sphere radius \(r_{\rm ph}\) and the shadow radius \(R_{\rm sh}\) grow monotonically. Although these enlarged shadows are compatible with either regular or singular end states, the matching procedure between the interior and exterior solutions indicates that a phase transition occurs during collapse. Because this transition takes place as the apparent horizons are forming and may even delay horizon formation further investigation is necessary to: Identify the microphysical mechanisms driving the phase transition in the collapsing matter. Determine observable signatures of this transition, in particular any transient energy flux emitted at the moment of the phase change. In order to characterize the emitted flux, one must account for quantum effects during collapse and estimate both its luminosity and spectral properties. Addressing these questions in future work will clarify whether the ultimate fate of gravitational collapse is a truly regular black hole or still a singular configuration.}

\textcolor{black}{It is important to address the emission of electromagnetic radiation during gravitational collapse. As demonstrated in Ref.~\cite{Vertogradov:2025yto}, when baryonic matter undergoes a transition to a novel, non-singular phase, a finite amount of energy is radiated away in the form of electromagnetic waves. The energy density of this emitted radiation is determined by both the initial baryonic state and the properties of the final, regular matter phase. Consequently, the spectrum and total luminosity depend sensitively on the EoS parameters governing the transition. In particular, stronger deviations from the standard baryonic EoS or more rapid phase changes lead to more intense electromagnetic output, thereby providing a potential observational signature of singularity resolution in collapsing compact objects.}


\textcolor{black}{The gravitational collapse proceeds in three distinct stages. In the first stage, the entire star is composed of baryonic matter. As collapse advances and densities increase toward the center, baryonic matter begins to convert into a novel phase that supports a regular core; this conversion is accompanied by the emission of electromagnetic radiation. In the final stage, all baryonic material has transitioned into the new phase, the electromagnetic emission ceases, and the resulting black hole features a non-singular central region. It should be emphasized that this radiation can only escape to an external observer if the formation of apparent horizons is sufficiently delayed; consequently, the observable emission lasts for a very brief interval.}

\acknowledgments
A. {\"O}. would like to acknowledge the contribution of the COST Action CA21106 - COSMIC WISPers in the Dark Universe: Theory, astrophysics and experiments (CosmicWISPers), the COST Action CA22113 - Fundamental challenges in theoretical physics (THEORY-CHALLENGES) and CA23130 - Bridging high and low energies in search of quantum gravity (BridgeQG). We also thank TUBITAK and SCOAP3 for their support.


\bibliographystyle{apsrev4-1}
\bibliography{references}
\end{document}